\documentclass[aps,preprint,nofootinbib,superscriptaddress,prc]{revtex4}
\usepackage{amssymb}
\usepackage{CJK,upgreek,fancyhdr}
\usepackage{epsfig}
\usepackage{ulem}
\usepackage[bookmarksnumbered,bookmarksopen,colorlinks,citecolor=blue,linkcolor=blue]{hyperref}
\begin{document}
\begin{CJK*}{GB}{gbsn}

\title{Statistical errors in Weizs\"acker-Skyrme mass model}

\author{Min Liu(ÁõÃô), Yu Gao(¸ßÓî), Ning Wang(ÍõÄþ)}

\thanks{wangning@gxnu.edu.cn}
\affiliation{Department of Physics, Guangxi Normal University,
Guilin 541004, P. R. China}

\affiliation{Guangxi Key Laboratory Breeding Base of Nuclear Physics and Technology,
Guilin 541004, P. R. China}

\begin{abstract}
  The statistical uncertainties of 13 model parameters in the Weizs\"acker-Skyrme (WS*) mass model are investigated for the first time with an efficient approach, and the propagated errors in the predicted masses are estimated. The discrepancies between the predicted masses and the experimental data, including the new data in AME2016, are almost all smaller than the model errors. For neutron-rich heavy nuclei, the model errors increase considerably, and go up to a few MeV when the nucleus approaches the neutron drip line. The most sensitive model parameter which causes the largest statistical error is analyzed for all bound nuclei. We find that the two coefficients of symmetry energy term significantly influence the mass predictions of extremely neutron-rich nuclei, and the deformation energy coefficients play a key role for well-deformed nuclei around the $\beta$-stability line.

\end{abstract}

\maketitle

\begin{center}
\textbf{I. INTRODUCTION}
\end{center}

Nuclear masses, as one of the basic quantities in nuclear physics, play crucial roles not only in the study of nuclear structure and reactions, but also in the study of astrophysics, such as understanding the origin of elements in the universe. The nuclear mass models \cite{Lun03,Moll95,HFB17,HFB27,Zhao10,Geng05,Meng13,Zhao1,GK,Brown,SunY,DZ28,Wang,Wang10,Liu11,Wang14}, including global and local mass models, are of significant importance for exploring the exotic structure of extremely neutron-rich nuclei, as well as the structures of super-heavy nuclei and their decay properties \cite{Naza,Sob,Sob16,Ogan15, Zhou12,WangYZ}. In addition, the nuclear mass models are also helpful in the investigation of nuclear symmetry energy \cite{Dani14,LiBA,Liu10,Jiang15,WangN,FRDM12}, which probes the isospin part of nuclear forces, because the symmetry energy coefficient in nuclear mass models significantly affects the masses of heavy nuclei near the neutron drip line.

Up to now, a number of nuclear mass models have been developed with root-mean-square (rms) deviations of about several hundred keV to one MeV with respect to all known masses. For example, a macroscopic-microscopic mass model, the Weizs\"acker-Skyrme (WS*) model \cite{Wang10}, which is inspired by the Skyrme energy-density functional and the isospin symmetry of nuclear force, was proposed with an RMS deviation of 441 keV with respect to the 2149 measured masses \cite{Audi} in the 2003 Atomic Mass Evaluation (AME2003). For unmeasured nuclear masses, the discrepancies between different model predictions are still large and even larger than 20 MeV for heavy nuclei near the neutron drip line \cite{Wang15}. It is therefore important and interesting to estimate the uncertainties of mass predictions and the predictive power of these different mass models. Unfortunately, it is difficult to accurately calculate the uncertainties of model predictions due to the complicated parameter space and limited computational power, and thus most nuclear mass models omit the theoretical estimation of errors and correlations between parameters. In recent years, estimates of extrapolation errors of theoretical models from different strategies such as least-squares fit, covariance analysis, variation of fit data, and so on, have attracted a lot of attention \cite{Gao13,Erler15,Maza15,Yuan16}. Covariance analysis is a useful tool for understanding the limitations of a model, the correlations between observables and the statistical errors, with which the statistical errors in the parameters of nuclear energy density functionals and in some predicted observables such as neutron-skin thickness of $^{208}$Pb are investigated \cite{Maza15}. Although the statistical errors in the parameters of some energy-density functionals have been studied in the literature \cite{Kort15,Schunck,Nik15}, a systematic study of statistical errors in the predicted masses of all bound nuclei, especially the unmeasured extremely neutron-rich nuclei and super-heavy nuclei, has not yet been performed based on the macroscopic-microscopic mass models. In addition, it is interesting to investigate the influence of parameter sensitivity on the uncertainty of predicted masses of extremely neutron-rich nuclei and super-heavy nuclei.

In this work, we attempt to study the statistical uncertainties in the 13 parameters of the WS* mass model \cite{Wang10} and the corresponding model errors in mass predictions, with a more efficient approach rather than the traditional covariance matrix method. The paper is organized as follows. In Section 2, the WS* mass model is briefly introduced. In Section 3, the procedure of extraction of the statistical uncertainties in parameters and the estimation of the model errors will be introduced. The calculated results will also be presented. Finally, a summary is given in Section 4.

\begin{center}
\textbf{II. THE WS* NUCLEAR MASS MODEL}
\end{center}

The WS* nuclear mass model is based on the macroscopic-microscopic method. The total
energy of a nucleus is expressed as a sum of the liquid-drop
energy and the Strutinsky shell correction $\Delta E$,
\begin{eqnarray}
E (A,Z,\beta)=E_{\rm LD}(A,Z) \prod_{k \ge 2} \left (1+b_k
\beta_k^2 \right )+\Delta E (A,Z,\beta).
\end{eqnarray}
The liquid drop energy of a spherical nucleus $E_{\rm LD}(A,Z)$ is
described by a modified Bethe-Weizs\"acker mass formula,
\begin{eqnarray}
E_{\rm LD}(A,Z)=a_{v} A + a_{s} A^{2/3}+ E_C + a_{\rm sym} I^2 A +
a_{\rm pair}  A^{-1/3}\delta_{np}
\end{eqnarray}
with the Coulomb energy term,
\begin{eqnarray}
E_C=a_{c}\frac{Z^{2}}{A^{1/3}} \left [ 1- Z^{-2/3} \right].
\end{eqnarray}
$a_{\rm sym}$ is the symmetry energy
coefficient with isospin asymmetry $I=(N-Z)/A$,
\begin{eqnarray}
 a_{\rm sym}=c_{\rm sym}\left [1-\frac{\kappa}{A^{1/3}}+\frac{2-|I|}{ 2+|I|A} \ \right
 ].
\end{eqnarray}
The terms with $b_k$ describe the contribution of nuclear deformation to
the macroscopic energy, which is efficient in sharply reducing the CPU hours needed in the calculations of deformed nuclei,
\begin{eqnarray}
b_k=\left ( \frac{k}{2} \right ) g_1A^{1/3}+\left ( \frac{k}{2}
\right )^2 g_2 A^{-1/3}.
\end{eqnarray}

The microscopic shell correction of a nucleus, obtained by the traditional Strutinsky
procedure, is:
\begin{eqnarray}
\Delta E=c_1 E_{\rm sh} + |I| E_{\rm sh}^{\prime}.
\end{eqnarray}
Here, $E_{\rm sh}$ and $E_{\rm sh}^{\prime}$ denote the shell
energy of a nucleus and of its mirror nucleus, respectively. The
additionally introduced $|I| E_{\rm sh}^{\prime}$ term is to
take into account the mirror constraint from the
isospin symmetry. In the calculations of shell corrections, the single particle levels of a nucleus are calculated under the axially deformed Woods-Saxon potential with four parameters: depth of the potential $V_0$, radius coefficient of the potential $r_0$, surface diffuseness $a$, and strength of the spin-orbit potential $\lambda_0$.

\begin{table}
\caption{ RMS deviations between data ($N \ge 8$, $Z \ge 8$) and model
predictions from different mass models (in keV). The row $\sigma(M)$ refers to all the 2149 measured masses in AME2003, and the
row $\sigma (M_{\rm new})$ to the measured masses of 270 ``new" nuclei in AME2016 \cite{AME2016}.  }
\begin{tabular}{cccc}
 \hline\hline
                          & ~~~~WS*~~  & ~~FRDM\cite{Moll95}~~ & DZ28\cite{DZ28} \\
\hline
 $\sigma  (M)$           & $441$ & $656$ &  $360 $\\
 $\sigma  (M_{\rm new})$ & $589$ & $901$ &  $763 $\\
\hline\hline
\end{tabular}
\end{table}

In Table I we list the RMS deviations $\sigma (M)$ between experimental masses and the
predictions of some models (in keV). The RMS deviations with respect to
the data in AME2003 are 441, 656 and 360 keV from the predictions of WS*, finite range droplet model (FRDM)  and Duflo-Zuker (DZ28) model, respectively. The optimal values of the parameters in the three models listed in Table I are mainly determined by the measured masses in AME2003 or earlier data. Very recently, the latest atomic mass evaluation table AME2016 was published \cite{AME2016}, in which the measured masses of 270 new nuclei (since AME2003) are presented. These newly measured unstable nuclei are extremely neutron-rich or neutron-deficient, which is very helpful to test the predictive power of theoretical models. The RMS deviations of the three models with respect to the data of the 270 new nuclei in AME2016 go up to 589, 901 and 763 keV, respectively. The result of the WS* model is the best in the description of the masses of these new nuclei. Considering that the number of model parameters is only 13 in the WS* model, which is much smaller than the two other models, the WS* model provides us with a useful balance between accuracy and computation cost in performing a systematic study of the statistical errors in mass predictions.

\begin{center}
\textbf{III. STATISTICAL UNCERTAINTIES IN MODEL PARAMETERS AND MASS PREDICTIONS }
\end{center}

\begin{table}
\caption{ Optimal values and statistical uncertainties of model parameters in the WS* mass model. }
\begin{tabular}{ccc }
\hline\hline
  parameter          & ~~~~~~~WS*~~~~~~~      &  ~~~~$\sigma_i$~~~~\\ \hline
 $a_v  \; $ (MeV)            &   $-15.6223$  & 0.0030 \\
 $a_s \; $  (MeV)            &   18.0571     & 0.0156 \\
 $a_c \; $ (MeV)             &   0.7194      & 0.0007 \\
 $c_{\rm sym} $(MeV)         &   29.1563     & 0.1298 \\
 $\kappa \;  $               &   1.3484      & 0.0186 \\
 $a_{\rm pair} $(MeV)        &   $-5.4423$   & 1.9095 \\
 $g_1 $                      &   0.00895     & 0.0024 \\
 $g_2 $                      &   $-0.4632$   & 0.0668 \\
 $c_1  \; $                  &   0.6297      & 0.1566 \\
 $V_0$ (MeV)                 &   $-46.8784$  & 3.3602 \\
 $r_0$ (fm)                  &   1.3840      & 0.0835 \\
 $a $ (fm)                   &   0.7842      & 0.1038 \\
 $\lambda_0$                 &   26.3163     & 2.7898 \\

 \hline\hline
\end{tabular}
\end{table}

In the WS* mass model, there are 13 independent parameters, and the energy of a certain nucleus in its ground state is expressed as a function of these 13 model parameters: $E(a_v, a_s, a_c, c_{\rm sym}, \kappa,a_{\rm pair}, g_1, g_2, c_1, V_0, r_0, a, \lambda_0)$. The optimal values of these parameters are fixed by the measured masses of 2149 nuclei given in AME2003 and listed in Table 2. Here, the optimal values of the parameters are obtained from the masses of all 2149 nuclei rather than the mass of a certain nucleus, and for a certain nucleus the ``best" values of the parameters could be different from the optimal values listed in Table 2. In the traditional statistical error analysis to nuclear energy density functional \cite{Gao13}, the covariance matrix needs to be calculated.
The calculation of the covariance matrix is time consuming due to the huge number of all bound nuclei and the complicated parameter space. In this work, we attempt to analyze the statistical error of the macroscopic-microscopic mass model in a more efficient way. In the macroscopic-microscopic mass models, the correlations between the model parameters of the macroscopic part and those of the microscopic part are weak, which may provide us with an opportunity to investigate the statistical uncertainties in the model parameters independently.

\begin{figure}
\includegraphics[angle=-0,width=1.0\textwidth]{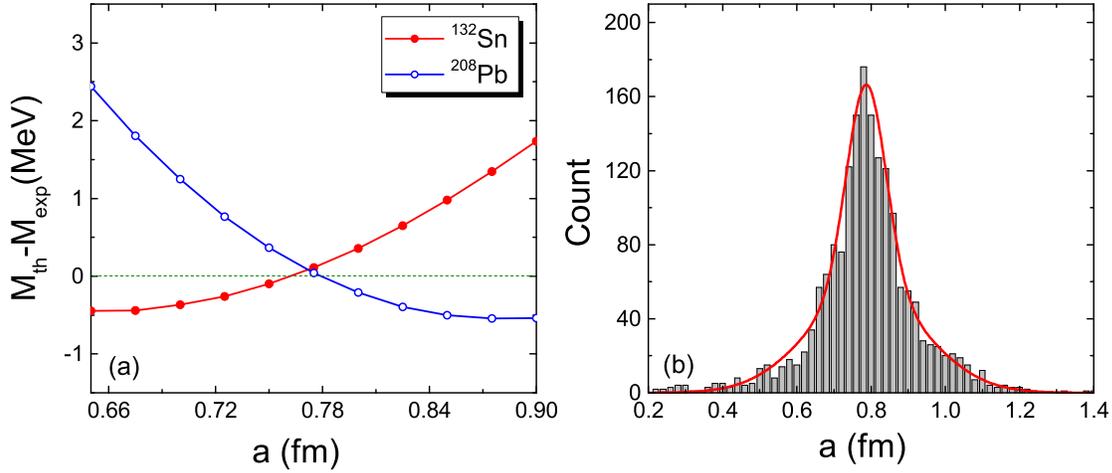}
 \caption{(Color online)  (a) Discrepancies between the experimental data and  the calculated masses for $^{132}$Sn and $^{208}$Pb with the WS* model as a function of diffuseness parameter $a$ of the Woods-Saxon potential.  (b) Distribution of the ``best" value of $a$ from all 2149 measured nuclei. The solid curve in (b) denotes the Gaussian fit to the distribution. }
\end{figure}

\begin{figure}
\includegraphics[angle=-0,width=1.0\textwidth]{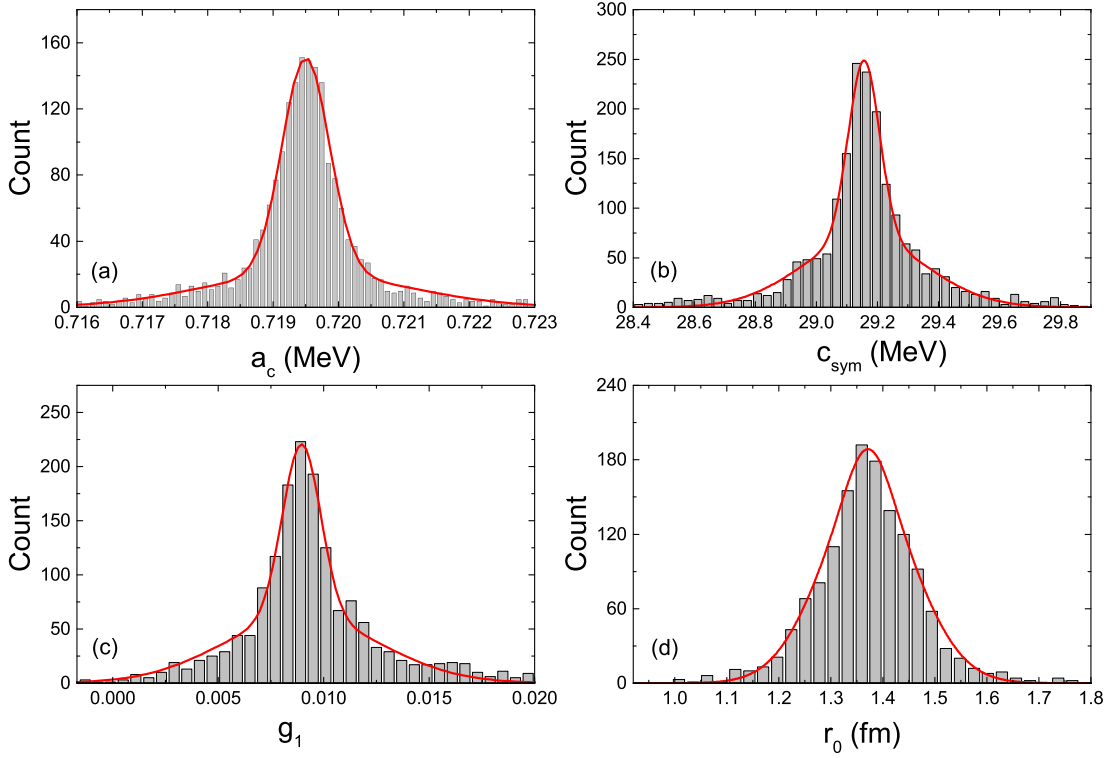}
 \caption{(Color online) Distributions of Coulomb energy coefficient $a_c$, volume symmetry energy coefficient $c_{\rm sym}$, deformation energy coefficient $g_1$, and potential radius coefficient $r_0$. The solid curve denotes the Gaussian fit to the distribution.  }
\end{figure}

In this work, the statistical errors in the model parameters are obtained based on maximum likelihood estimation. More specifically, for a certain model parameter, e.g. the diffuseness parameter $a$ of the Woods-Saxon potential, we calculate the energy $E$ of a certain nucleus by varying the value of this parameter around its optimal value and keeping other parameters unchanged. If the discrepancy of the calculated mass for a certain nucleus from the corresponding experimental data equals zero, the ``best" value of this parameter for the given nucleus is therefore obtained. In Fig. 1(a) we show, as an example, the discrepancies of the calculated masses for $^{132}$Sn and $^{208}$Pb from the experimental data as a function of the diffuseness parameter $a$. The ``best" values of $a$ for $^{132}$Sn and $^{208}$Pb are slightly different from each other. For all the measured nuclei, we can obtain a distribution of the ``best" value of $a$. The peak of the distribution is generally located at the optimal value $a=0.7842$ fm given in the WS* model. With the same procedure, the distribution of the other 12 parameters can also be obtained. Some results are shown in Fig. 2. We find that the obtained distributions can be reasonably well described by using two Gaussian functions with the same centroid but different widths. The shoulder and long tail of the distribution may come from the influence of other parameters.
Based on the obtained distributions of the model parameters, the statistical uncertainties in the 13 parameters can be extracted with 68.3\% confidence level. In Table 2, we also list the standard deviation $\sigma_i$ for each model parameter. We note that the statistical uncertainties in the pairing coefficient $a_{\rm pair}$, the deformation energy coefficient $g_1$, the shell correction factor $c_1$, the depth of the Woods-Saxon potential $V_0$, and the strength of spin-orbit potential $\lambda_0$ are relatively large.

\begin{figure}
\includegraphics[angle=-0,width=1.0\textwidth]{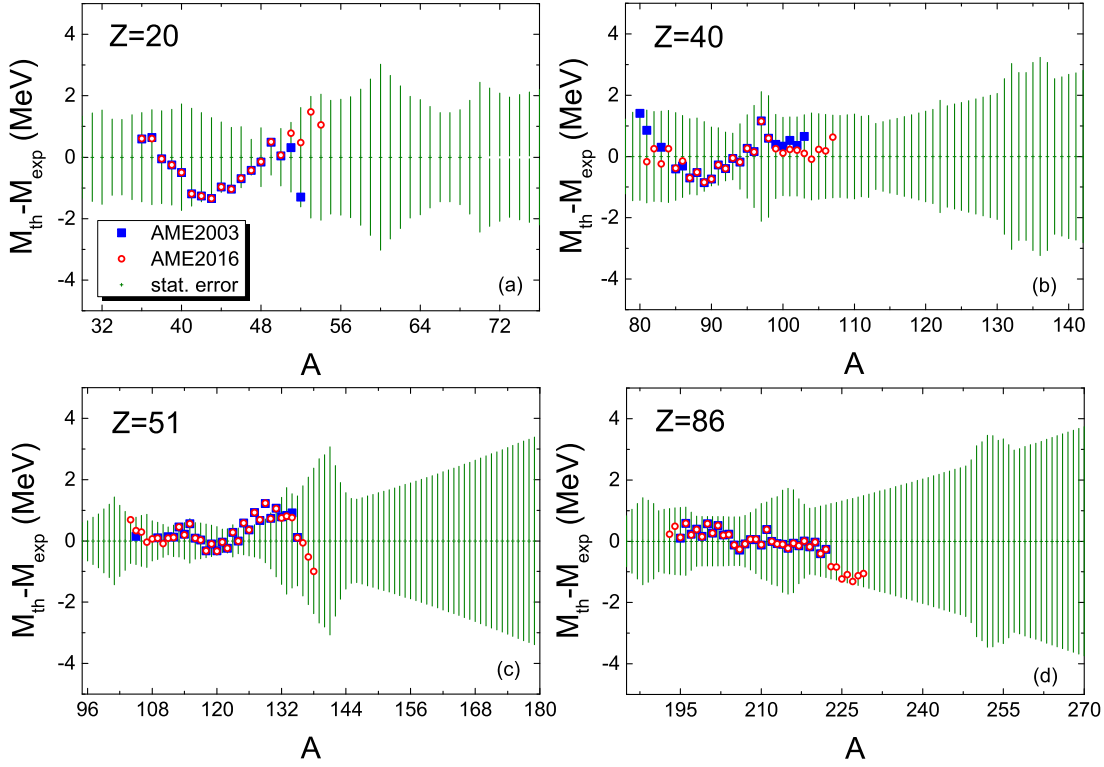}
 \caption{(Color online)  Difference between the measured masses of Ca, Zr, Sb, Rn isotopes and the predicted results from the WS* model. The squares and circles denote the data from AME2003 and AME2016, respectively. The error bars denote the statistical errors in the predicted masses. }
\end{figure}

Based on the extracted statistical uncertainties $\sigma_i$ in the model parameters, we further investigate the statistical error in the predicted masses induced by the uncertainties in the parameters. In this work, the statistical error $\delta E$ in the predicted energy for a certain nucleus at its ground state is estimated by the maximal energy uncertainty considering the cancellation from different parameters, i.e.,
\begin{eqnarray}
\delta E=\max \left (\delta E_1, \cdots, \delta E_i, \cdots, \delta E_{13} \right ).
\end{eqnarray}
Here, $\delta E_i$ denotes the uncertainty of the ground state energy of a nucleus induced by the uncertainty $\sigma_i$ of the $i$-th model parameter $x_i$,
\begin{eqnarray}
\delta E_i=\left |E(x_1,\cdots, x_i+\sigma_i, \cdots, x_{13})- E(x_1,\cdots, x_i, \cdots, x_{13}) \right |.
\end{eqnarray}
In Fig. 3, we show the discrepancies between the experimental data and the calculated masses for Ca, Zr, Sb, Rn isotopes  using the WS* model. The squares and circles denote the data from AME2003 and AME2016, respectively. The error bars denote the statistical errors $\delta E$ in the predicted masses according to Eq.~(7). One can see from the figure that $\delta E$ is different for different nuclei. The discrepancies for all these nuclei, not only the nuclei in AME2003 (squares) but also the new data in AME2016 (circles), are almost all located in the range of the error bars, which indicates that the proposed estimation for the statistical error in the predicted masses is reasonable. For extremely neutron-rich heavy nuclei, $\delta E$ clearly increases with neutron number, which is mainly due to the symmetry energy term. In addition, the uncertainties do not increase monotonically  in some mass regions, which is due to the competition among different model parameters. For example, the uncertainty in the symmetry energy is proportional to $I^2 A$ whereas the uncertainty in the deformation energy is proportional to nuclear deformations. To see the global behavior of the statistical errors, we show in Fig. 4 the values of $\delta E$ for almost all nuclei in the nuclear landscape. For intermediate and heavy nuclei around the $\beta$-stability line, the statistical errors in the predicted masses are generally smaller than 1 MeV. For super-heavy nuclei with neutron number larger than 180 and heavy nuclei approaching the neutron drip line, the statistical errors increase significantly, even larger than 4 MeV.

\begin{figure}
\includegraphics[angle=-0,width=0.7\textwidth]{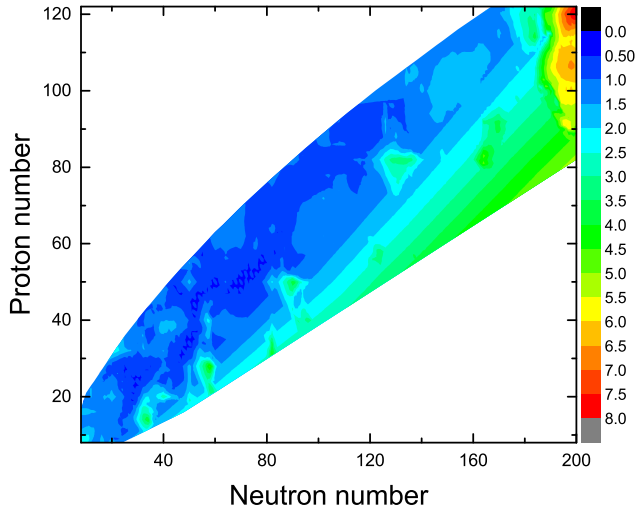}
 \caption{(Color online) Statistical errors $\delta E$ in the predicted masses for almost all nuclei in the nuclear landscape.  }
\end{figure}

To understand the influence of the model parameters on the mass predictions, we simultaneously investigate the most sensitive parameter in the mass calculations for a certain nucleus. Here, the most sensitive parameter means the parameter that results in the largest statistical uncertainty among the 13 $\delta E_i$ from Eq.~(8). In Fig. 5, we show the distribution of the most sensitive parameters for almost all nuclei in the nuclear landscape. We find that for nuclei approaching the neutron drip line, the volume symmetry energy coefficient $c_{\rm sym}$ and surface-symmetry coefficient $\kappa$ play a key role in the mass predictions. For extremely neutron-rich intermediate-mass nuclei, the influence of the surface-symmetry energy term is relatively stronger, due to  the mass dependence of the symmetry energy coefficient $a_{\rm sym}$ becoming stronger in the intermediate-mass region compared with the heavy-mass region. For well-deformed nuclei around the $\beta$-stability line, the deformation energy coefficients $g_1$, $g_2$ play a role in accurate mass predictions. In addition, the radius parameter $r_0$ of the single-particle Woods-Saxon potential and the strength $\lambda_0$ of the spin-orbit potential strongly influence the masses of light nuclei and nearly-spherical nuclei.

\begin{figure}
\includegraphics[angle=-0,width=0.7\textwidth]{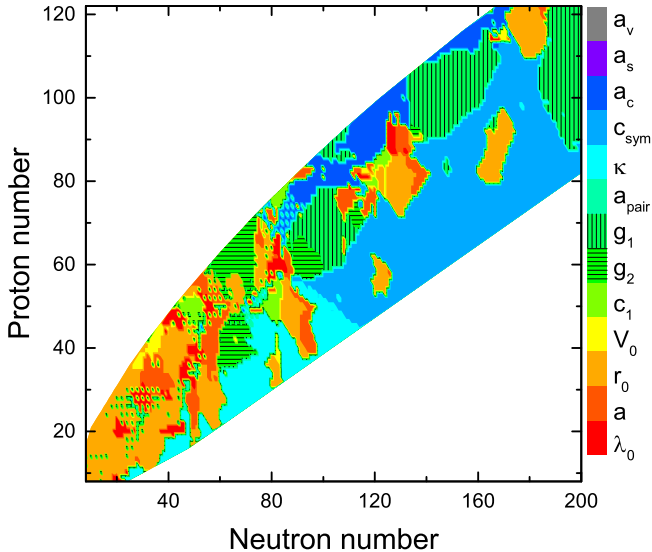}
 \caption{(Color online) Distribution of the most sensitive parameter.   }
\end{figure}

\begin{center}
\textbf{IV. SUMMARY AND DISCUSSION}
\end{center}

In this work, the statistical uncertainties in the 13 model parameters of the Weizs\"acker-Skyrme (WS*) mass model are investigated, and at the same time the propagated statistical errors in the predicted masses of measured and unmeasured nuclei are estimated with an efficient approach considering the weak correlations between the parameters of the macroscopic part and those of the microscopic part. The RMS deviations with respect to the masses of 270 new nuclei in AME2016 is only 589 keV from the WS* model, with 13 independent model parameters. By varying the value of one parameter around its optimal value given in the WS* model and checking the discrepancy between the experimental data and model prediction for a certain nucleus, one can obtain the ``best" value of this parameter for a given nucleus if the discrepancy equals zero. The statistical distribution of the ``best" values of the parameter is finally obtained according to all measured masses. The statistical uncertainties in the pairing coefficient $a_{\rm pair}$, the deformation energy coefficient $g_1$, the shell correction factor $c_1$, the depth of the Woods-Saxon potential $V_0$ and the strength of spin-orbit potential $\lambda_0$ are relatively large. The statistical error $\delta E$ in the predicted energy of a certain nucleus in its ground state is estimated by the maximal energy uncertainty due to the statistical uncertainties in the 13 parameters, which is tested by the differences between the predicted masses and the experimental data in AME2003 and the new data in AME2016. The discrepancies from the data are almost all smaller than the statistical errors estimated with the proposed approach. In addition, we investigate the most sensitive parameter in the mass calculations for a certain nucleus, and find that for nuclei approaching the neutron drip line, the volume symmetry energy coefficient $c_{\rm sym}$ and surface-symmetry coefficient $\kappa$ play a key role in the mass predictions.

In the proposed approach, we assume that the correlations between the model parameters of WS* are weak in the estimation of the model statistical errors. To check the assumption, the correlations between any two parameters in the WS* model are also studied and the Pearson correlation coefficient $r$ is calculated. We find that the correlations between the parameters of the macroscopic part and those of the microscopic part is very weak, as expected, and the corresponding absolute values of $r$ are smaller than 0.3. The correlations between the parameters of the macroscopic part are relatively strong, with $0.5\lesssim|r|\lesssim0.8$ in general, and the correlations between the parameters of the Woods-Saxon potential are also weak, with $|r|<0.6$. If the correlations between the parameters of the macroscopic part are considered, the statistical error in the predicted masses could be slightly reduced.

\begin{center}
\textbf{ACKNOWLEDGEMENTS}
\end{center}

This work was supported by National Natural Science Foundation of
China (Nos.11422548, 11365005 and 11365004) and Guangxi Natural Science Foundation (No. 2015GXNSFDA139004).  We thank Liyong Liang for performing the mass calculations with the WS* model. The WS* mass table with the statistical errors obtained by using the proposed approach is available at http://www.imqmd.com/mass/.

\clearpage
\end{CJK*}
\end{document}